\documentclass[a4paper]{jpconf}
\usepackage{graphicx}
\usepackage{subfigure}
\usepackage{amsmath,amssymb}

\makeatother

\newcommand{\bra}[1]{\left\langle #1 \right|}

\newcommand{\ket}[1]{\left| #1 \right\rangle}

\newcommand{\Hhat}{\hat{H}}

\newcommand{\Jc}{{\cal J}}

\renewcommand{\b}{\beta}
\newcommand{\g}{\gamma}

\newcommand{\bg}{\beta,\gamma}
\newcommand{\kr}[1]{$^{#1}$Kr}

\newcommand{\Qhat}{\hat{Q}}

\newcommand{\Phat}{\hat{P}}

\newcommand{\beq}{\begin{equation}}
\newcommand{\beqa}{\begin{eqnarray}}
\newcommand{\eeq}{\end{equation}}
\newcommand{\eeqa}{\end{eqnarray}}

\renewcommand{\thanks}{\footnote}

\begin{document}
\title{Microscopic approach to large-amplitude deformation dynamics with local QRPA inertial masses}

\author{Koichi Sato$^{1}$, Nobuo Hinohara$^{1}$
, Takashi Nakatsukasa$^{1}$, Masayuki Matsuo$^{2}$, Kenichi Matsuyanagi$^{1,3}$
}

\address{
$^{1}$RIKEN Nishina Center, Wako 351-0198, Japan \\
$^{2}$Department of Physics,
Niigata University, Niigata 950-2181, Japan\\
$^{3}$Yukawa Institute for Theoretical Physics, Kyoto University, 
Kyoto 606-8502, Japan
}

\ead{satok@ribf.riken.jp}

\begin{abstract}
We have developed a new method for determining microscopically 
the five-dimensional quadrupole collective Hamiltonian, 
on the basis of the adiabatic self-consistent collective coordinate method. 
This method consists of the constrained Hartree-Fock-Bogoliubov (HFB) equation and the local QRPA (LQRPA) equations, 
which are an extension of the usual QRPA (quasiparticle random phase approximation) 
to non-HFB-equilibrium points, on top of the CHFB states. 
One of the advantages of our method is that the inertial functions calculated with this method 
contain the contributions of the time-odd components of the mean field, 
which are ignored in the widely-used cranking formula.
We illustrate usefulness of our method by applying to 
oblate-prolate shape coexistence in $^{72}$Kr and 
shape phase transition in neutron-rich Cr isotopes around $N=40$. 
\end{abstract}

\section{Introduction}
%
%
The five-dimensional (5D) quadrupole collective Hamiltonian 
is a powerful tool to describe large-amplitude 
deformation dynamics including triaxial shape degree of freedom
as well as axially symmetric deformation. 
The 5D quadrupole collective Hamiltonian 
is characterized by seven functions: the collective potential, three vibrational and 
three rotational inertial functions (collective masses).
To calculate the inertial functions, the Inglis-Belyaev (IB) cranking formula \cite{Inglis1954, Beliaev1961} 
has been widely used.
However, it is well known that IB cranking masses do not contain
the contributions from the time-odd components of the mean field
and underestimate the collective masses.

In this presentation, we introduce a new method of determining microscopically
the collective potential and the inertial functions which can
overcome the shortcoming of the IB cranking formula.
This method is derived from the adiabatic self-consistent
collective coordinate (ASCC) method \cite{Matsuo2000,Hinohara2007}, which is based on the adiabatic time-dependent
Hartree-Fock-Bogoliubov (TDHFB) theory and, with which one can extract the collective subspace (collective manifold) 
from the large-dimensional TDHFB configuration space.
The method we employ in this study is the version of the ASCC method simplified
by assuming the one-to-one correspondence between the
collective manifold and the quadrupole deformation parameter space $(\beta,\gamma)$.
The main concept of this method is the local normal modes on top of 
the constrained Hartree-Fock-Bogoliubov (CHFB) 
states at every point of  the $(\beta,\gamma)$ plane.
In this method, we first solve the CHFB equations imposing the constraints on the 
proton and neutron numbers and deformation parameters $(\beta,\gamma)$.
Then, we solve local QRPA (quasiparticle random-phase approximation) 
equations, which are an extension of the usual QRPA 
to the non-HFB-equilibrium points.
Hereafter, we call the method ``the constrained HFB plus local QRPA (CHFB+LQRPA) method.'' 

We shall show some results of the applications of the CHFB+LQRPA method 
to the oblate-prolate shape coexistence  in $^{72}$Kr
and to the shape phase transition in neutron-rich Cr isotopes around $N=40$.
%

\section{Theoretical framework}
Here, we briefly explain the theoretical framework of the CHFB+LQRPA method (see Ref. \cite{Hinohara2010} for details).
The 5D quadrupole collective Hamiltonian is written 
in terms of the magnitude $\beta$, the degree of triaxiality $\gamma$  
of quadrupole deformation, three Euler angles, 
and their time derivatives
as
\begin{align}
 \mathcal{H}_{\rm coll}&= T_{\rm vib} + T_{\rm rot} + V(\bg), \label{eq:Hc} \\
 T_{\rm vib}   &= \frac{1}{2} D_{\beta\beta}(\bg)\dot{\beta}^2 +
 D_{\beta\gamma}(\bg)\dot{\beta}\dot{\gamma}
+ \frac{1}{2}D_{\gamma\gamma}(\bg)\dot{\gamma}^2, \label{eq:Tvib}\\
 T_{\rm rot} &= \frac{1}{2}\sum_{k=1}^3 \Jc_k(\bg) \omega^2_k, \label{eq:collH_BM} 
\end{align}
where $T_{\rm vib}$, $T_{\rm rot}$ and $V$ represent  
the vibrational, rotational and collective potential energies, respectively.
We determine 
the three vibrational inertial masses, three rotational moments of inertia,
and collective potential in the collective Hamiltonian, by solving the CHFB+LQRPA equations.

In the CHFB+LQRPA method, we first solve the CHFB equation to determine the collective potential $V(\b,\g)$:   
\begin{align}
 \delta \bra{\phi(\bg)} \Hhat_{\rm CHFB}(\bg) \ket{\phi(\bg)} = 0,  \label{eq:CHFB} 
\end{align}
where $\Hhat_{\rm CHFB}=\Hhat-\sum_\tau\lambda^{(\tau)}\tilde N^{(\tau)}-\sum_m\mu^{(m)}\hat D_{2m}$.
Then, we solve the LQRPA equations on top of the CHFB states obtained above,
\begin{align}
 \delta & \bra{\phi(\bg)} [ \Hhat_{\rm CHFB}(\bg), \Qhat^i(\bg) ]
 - \frac{1}{i}\Phat_i(\bg) \ket{\phi(\bg)}  = 0, \label{eq:LQRPA1}\\
 \delta & \bra{\phi(\bg)} [ \Hhat_{\rm CHFB}(\bg), \frac{1}{i} \Phat_i(\bg)]
 - C_i(\bg) \Qhat^i(\bg) \ket{\phi(\bg)} = 0,  \quad\quad (i=1, 2). 
\label{eq:LQRPA2}
\end{align}
The vibrational inertial functions are calculated by transforming two LQRPA modes 
to the $(\beta, \gamma)$ degrees of freedom.
We also solve the LQRPA equations for rotation to determine the moments of inertia. 
In this work, we adopt a version of the pairing-plus-quadrupole (P+Q) interaction including the quadrupole 
pairing interaction as well as the monopole pairing interaction
and take two major harmonic oscillator shells as a model space.

Finally, we solve the collective Schr\"odinger equation for the
5D quadrupole collective Hamiltonian quantized according to Pauli's
prescription to obtain the excitation energies and collective wave functions,
with which the electric quadrupole transitions and moments are calculated.



\section{Application to oblate-prolate shape coexistence in $^{72}$Kr}

In this section, we briefly show the numerical results of the
application of the CHFB+LQRPA method to oblate-prolate shape coexistence 
in \kr{72}.
The effective charges are 
adjusted to $(e^{(n)}_{\rm eff},e^{(p)}_{\rm eff})= (0.658, 1.658)$ 
such that the calculated result of $B(E2;2_1^{+} \rightarrow 0_1^{+})$ reproduces the experimental value for $^{74}$Kr \cite{Clement2007}.  
The detailed results of this calculation are shown
in Ref. \cite{Sato2011}
together with the results for \kr{74,76}.

\begin{figure}[tb]
\begin{center}
\subfigure[$V(\beta,\gamma)$]{\includegraphics[height=0.35\textwidth,keepaspectratio,clip,trim=60 0 160 0]
{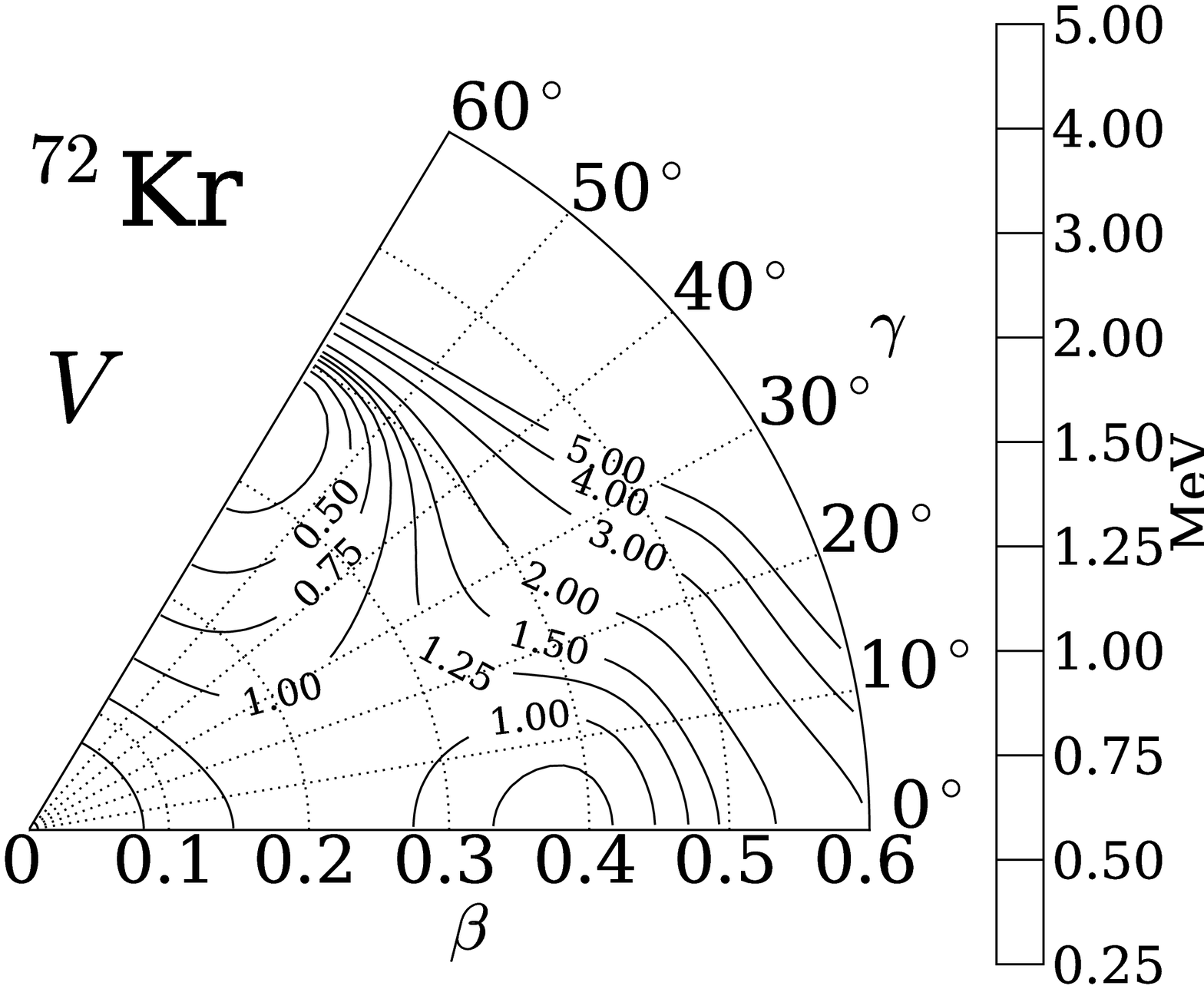}}
\subfigure[$D_{\b\b}(\bg)$]{
\includegraphics[height=0.35\textwidth,keepaspectratio,clip,trim=60 0 160 0]
{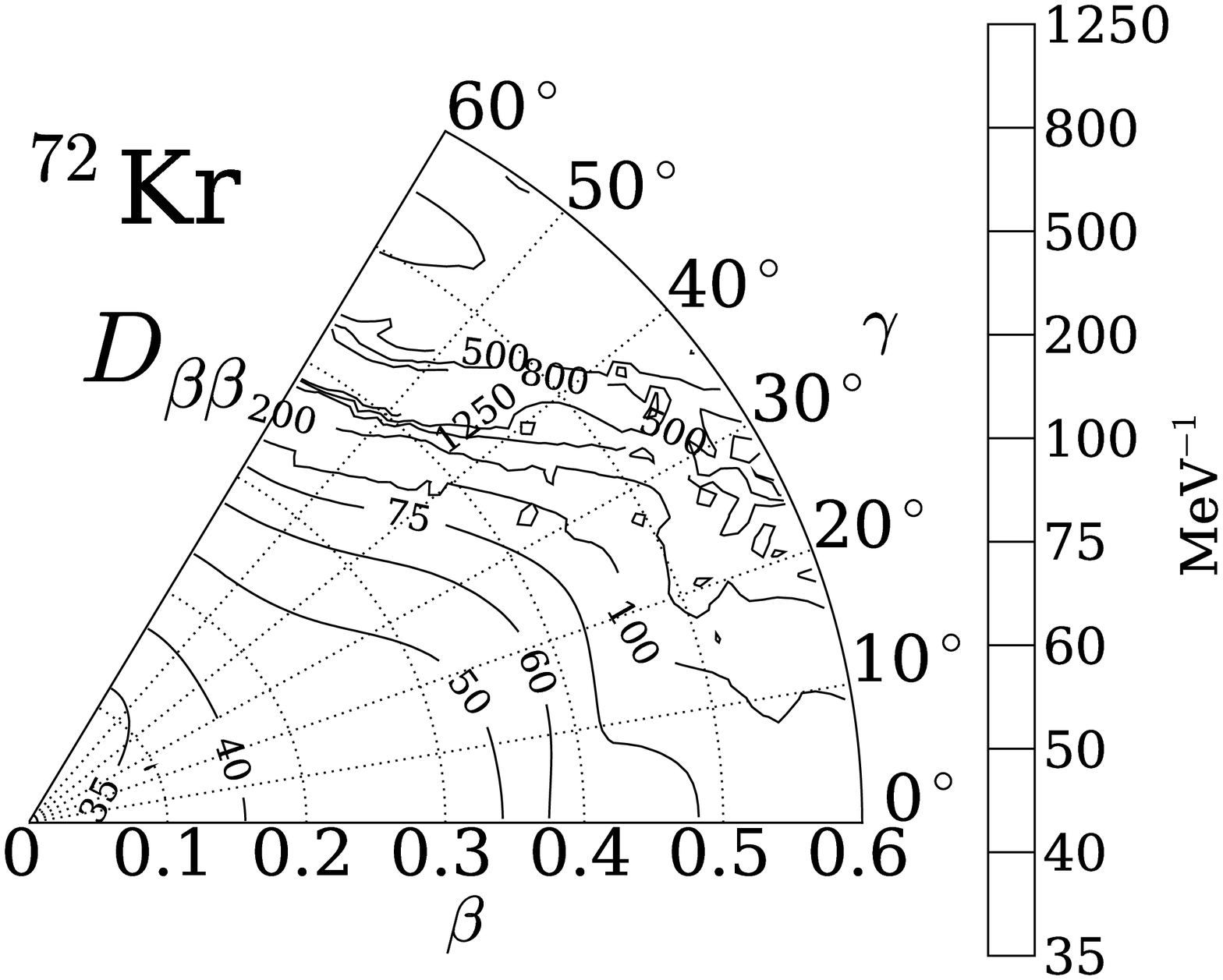}} 
\subfigure[$\Jc_{1}(\bg)$]{
\includegraphics[height=0.35\textwidth,keepaspectratio,clip,trim=20 0 160 0]
{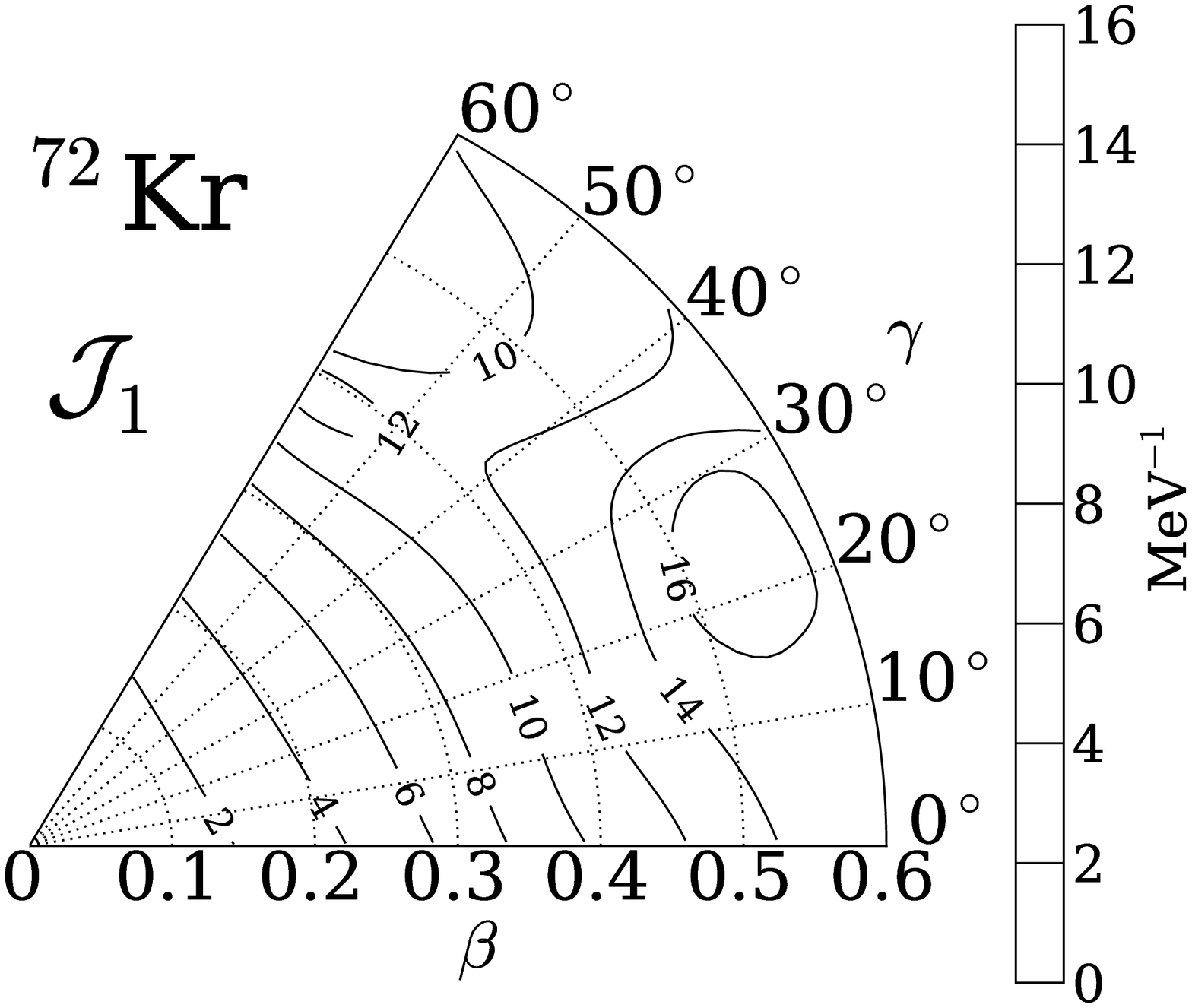}}
\end{center}
\caption{Collective potential energy surfaces $V(\beta,\gamma)$ in units of MeV, 
LQRPA vibrational inertial mass $D_{\beta\beta}(\bg)$ 
and LQRPA rotational moment of inertia $\Jc_{1}(\bg)$ 
in unit of MeV$^{-1}$
for $^{72}$Kr. }
\label{fig:VDJ}
\end{figure}

Figure \ref{fig:VDJ} shows the collective potential $V(\beta,\gamma)$, the vibrational inertial
function $D_{\beta\beta}(\beta,\gamma)$, and the rotational moment of inertia 
$\mathcal{J}_1(\beta,\gamma)$ for $^{72}$Kr.
The collective potential has two local minima: the oblate minimum is lower than the prolate one.
The spherical shape is a local maximum.
One can see that $D_{\b\b}$ indicates strong $\b-\g$ dependence 
and that the moment of inertia $\Jc_{1}$ deviates strongly from the irrotational moment of inertia. 
In Ref. \cite{Sato2011}, we have seen 
the collective wave function is localized on the oblate (prolate) side in the $(\beta,\gamma)$ plane in the ground
 (excited) band and that the localization of the wave function develops with increasing angular momentum. 
The development of the localization is strongly related to the $\beta-\gamma$ dependence of the moments of inertia 
seen in Fig. \ref{fig:VDJ}(c).

\begin{figure}[tb]
\begin{center}
\includegraphics[width=0.4\textwidth,angle=-90,keepaspectratio,clip]{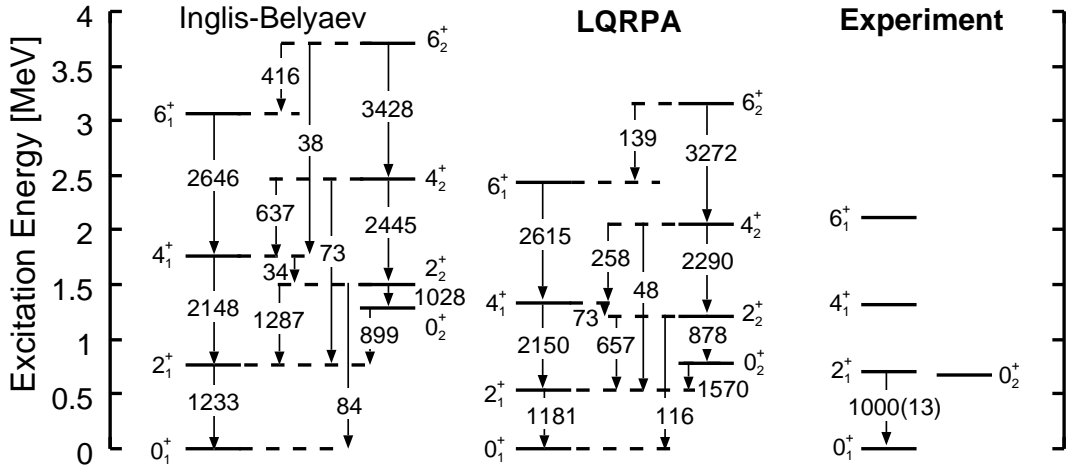}
\end{center}
\caption{Excitation spectra and $B(E2)$ values calculated for $^{72}$Kr 
by means of the CHFB+LQRPA method (denoted LQRPA) and 
experimental data  \cite{Bouchez2003,Gade2005,Fischer2003}.  
For comparison, results calculated using the IB cranking masses 
(denoted Inglis-Belyaev) are also shown. 
Only $B(E2)$'s larger than 1 Weisskopf unit  
are shown in units of $e^2{\rm fm}^4$.}
\label{fig:spectra72}
\end{figure}

In Fig. \ref{fig:spectra72}, the excitation energies and $B(E2)$ values for \kr{72} are shown together with the experimental data.
We show the result obtained using the IB cranking masses for comparison.
The excitation energies obtained with the LQRPA masses are lower than those obtained with the IB cranking masses
and agree better with the experimental data except for the $2_1^{+}$ state.
In particular, the observed $0_2^{+}$ excitation energy which is close to the $2_1^{+}$ excitation energy
is well reproduced with the LQRPA masses. 
One can see from the $E2$ transition strengths in Fig. \ref{fig:spectra72} that the shape-coexistence-like character becomes stronger with increasing angular momentum: the interband transitions between the initial and final states having equal angular momentum become weaker and weaker, 
which reflects the development of the localization of the vibrational wave function mentioned above.

%
%

\section{Application to shape transition in neutron-rich Cr isotopes}

In this section, we shall show the results of the application to
the CHFB+LQRPA method to shape phase transition in neutron-rich Cr isotopes 
around $N=40$, where the development of deformation has been suggested by recent
experimental data.
More details will be discussed in our forthcoming paper \cite{Sato2011b}.

\begin{figure}[tb]
\begin{center}
\subfigure[$^{58}$Cr]
{\includegraphics[height=0.35\textwidth,keepaspectratio,clip,trim=60 40 160 40]
{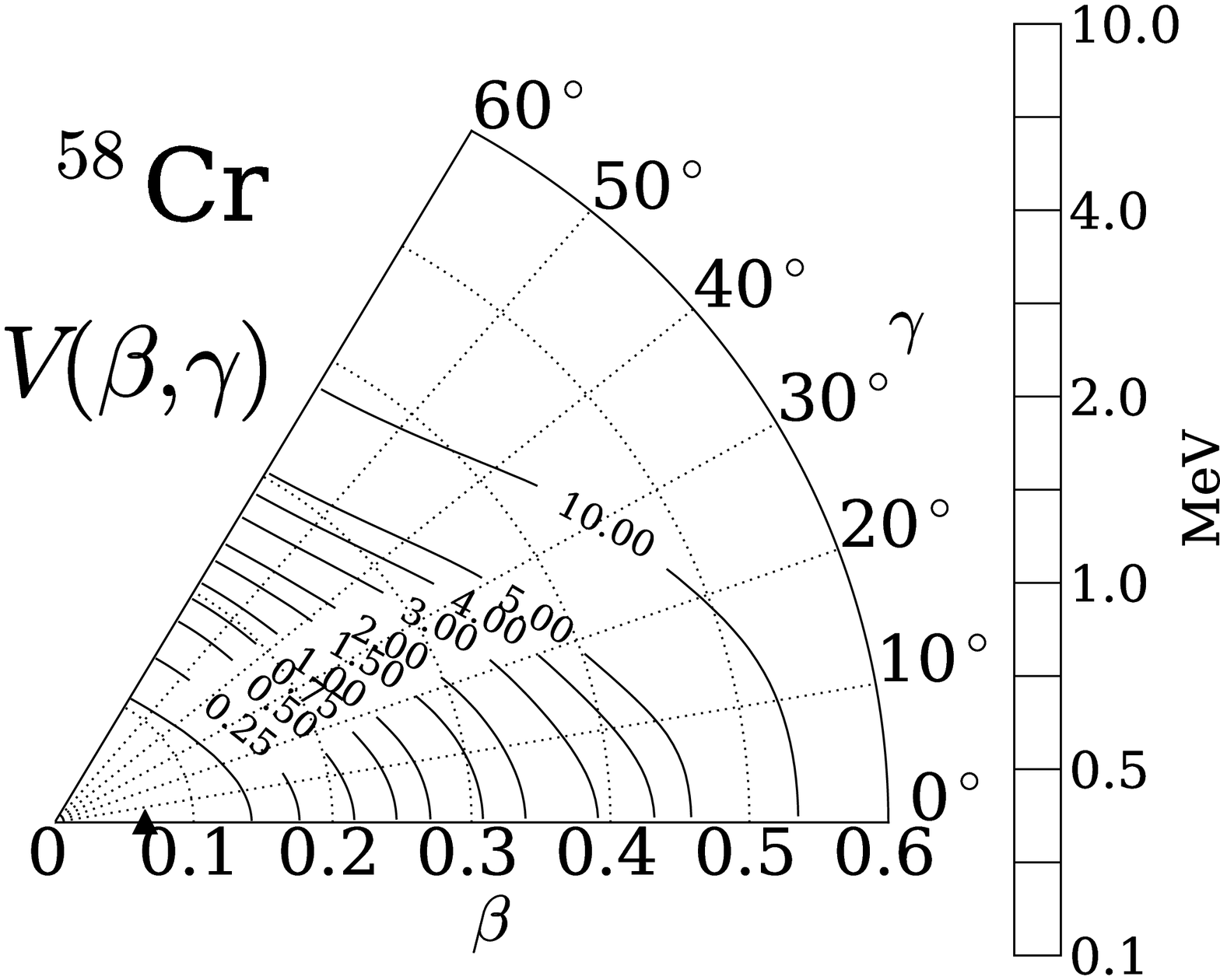}}
\subfigure[$^{60}$Cr]
{\includegraphics[height=0.35\textwidth,keepaspectratio,clip,trim=60 40 160 40]
{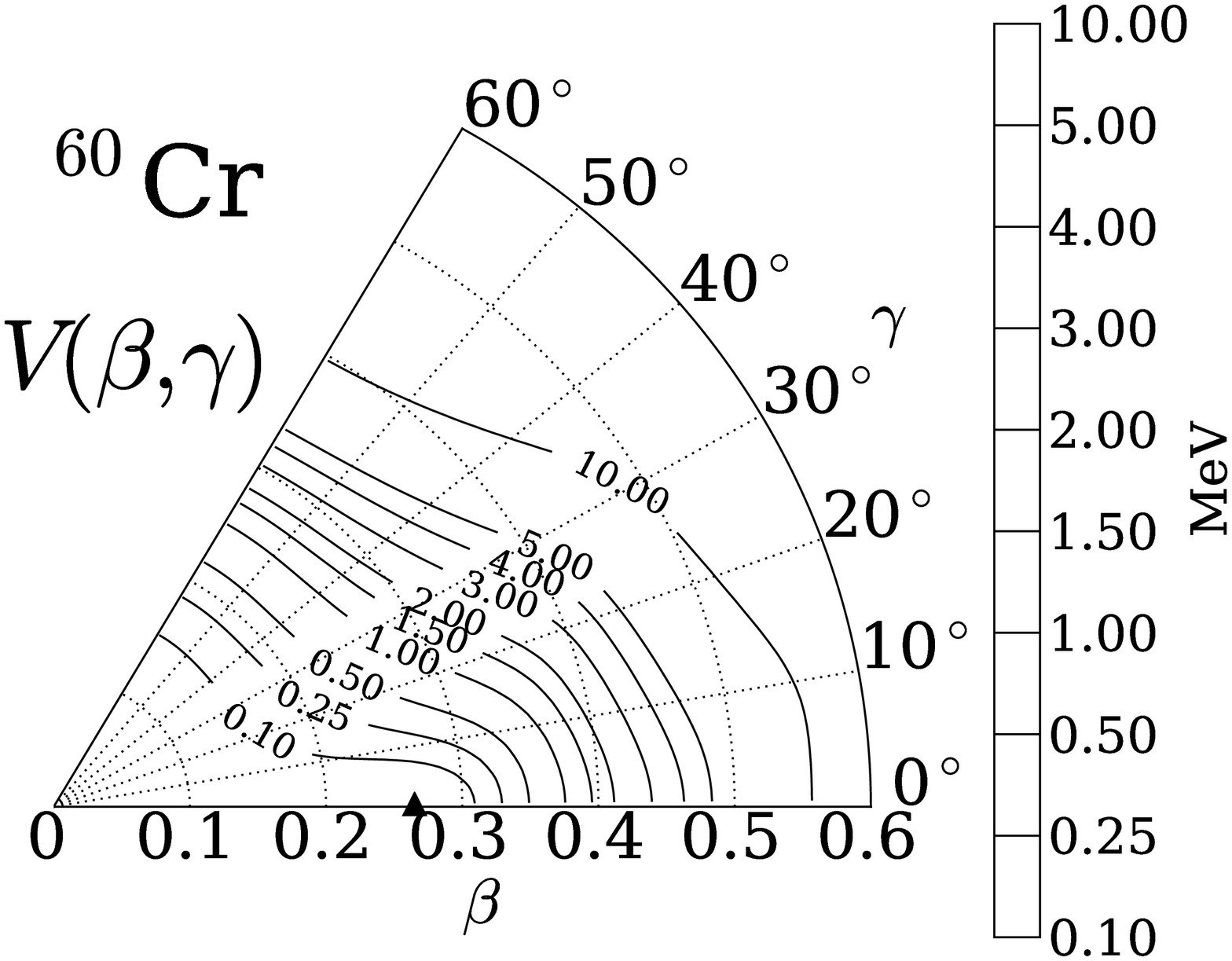}}
\subfigure[$^{62}$Cr]
{\includegraphics[height=0.35\textwidth,keepaspectratio,clip,trim=60 40 160 40]
{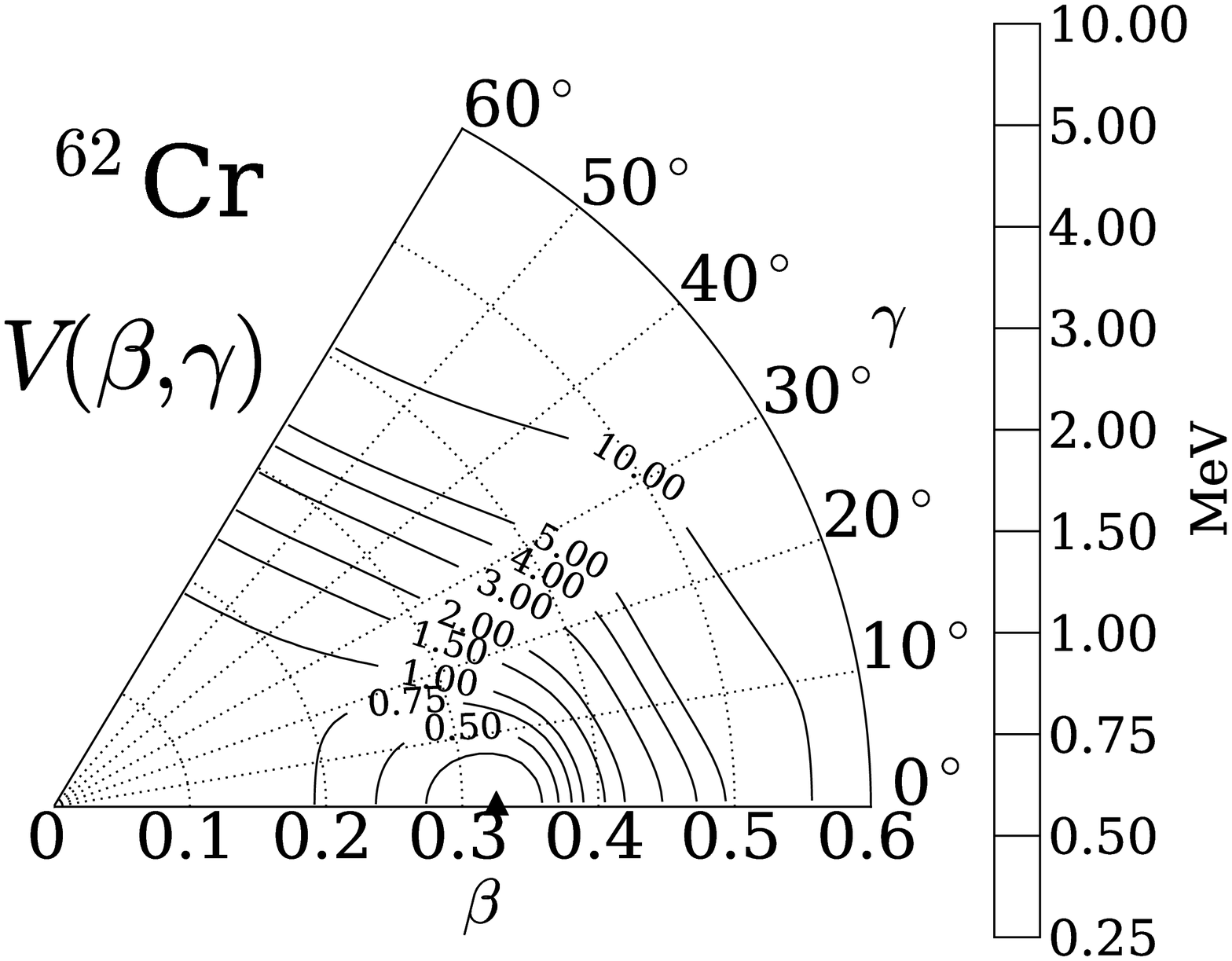}}
\subfigure[$^{64}$Cr]{
\includegraphics[height=0.35\textwidth,keepaspectratio,clip,trim=60 40 160 40]
{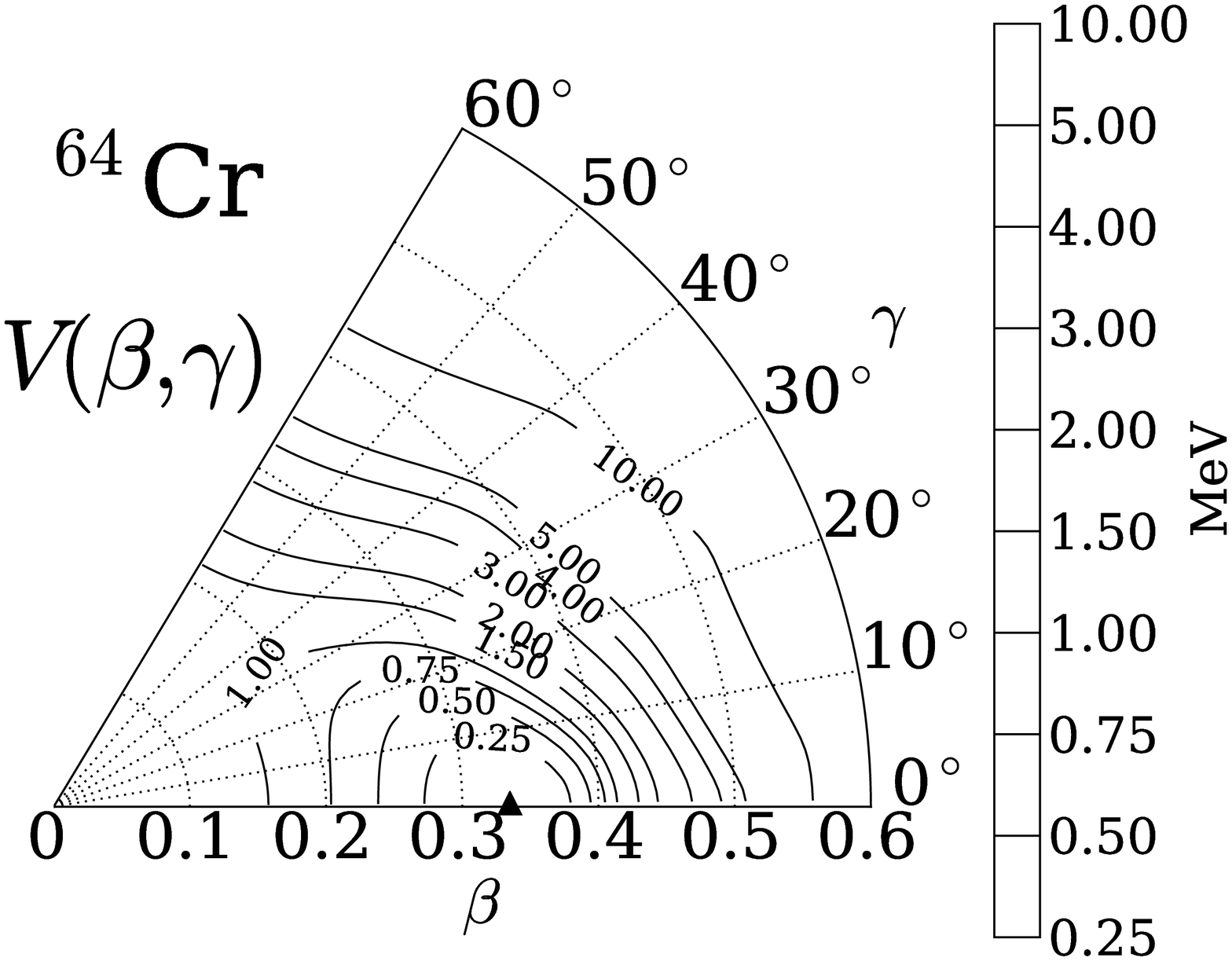}} 
\end{center}
\caption{Collective potential energy surfaces $V(\beta,\gamma)$ for $^{58-64}$Cr in units of MeV.
The triangle denotes the position of the absolute minimum.}
\label{fig:CrV}
\end{figure}

In this calculation, we have 
determined the interaction parameters as follows:
the parameters for $^{62}$Cr are determined such that
the calculated monopole pairing gaps and deformation at the HFB equilibrium point 
as well as the pairing gaps at the spherical shape reproduce those obtained with the
Skyrme-HFB calculation with the SkM* functional using the HFBTHO code \cite{Stoitsov2005}.
For the other nuclei $^{58,60,64,66}$Cr, we assume the simple mass number dependence according to 
Baranger and Kumar \cite{Baranger1968}.
We take two major harmonic oscillator shells with $N=3,4$ and $N=2,3$ for neutrons and protons, respectively.
The single particle energies are determined from those
obtained with the constrained Skyrme-HFB calculation at spherical shape. 
We scale them according to the effective mass of the SkM* functional $m^*/m=0.79$.
For the calculation of the $E2$ transitions and moments, we have used a standard value of 
the effective charges $(e^{(n)}_{\rm eff},e^{(p)}_{\rm eff})= (0.5, 1.5)$ .

We show the collective potentials for $^{58-64}$Cr in Fig. \ref{fig:CrV}.
The location of the absolute minimum is indicated by the triangle.
In $^{58}$Cr, the absolute minimum is located at a nearly spherical shape.
Although the minimum shifts to larger $\beta$ in $^{60}$Cr, the potential is extremely soft 
in the $\beta$ direction.
In $^{62}$Cr, 
a more definite local minimum appears
and the minimum gets still deeper in $^{64}$Cr. 
The potential energy surfaces seem to indicate
a shape transition from a spherical to a prolately deformed shape
along the isotopic chain toward $N=40$.

\begin{figure}[tb]
\begin{center}
\subfigure[$E(2_1^+)$]{\includegraphics[height=0.3\textwidth,keepaspectratio,clip,trim=0 0 0 0]
{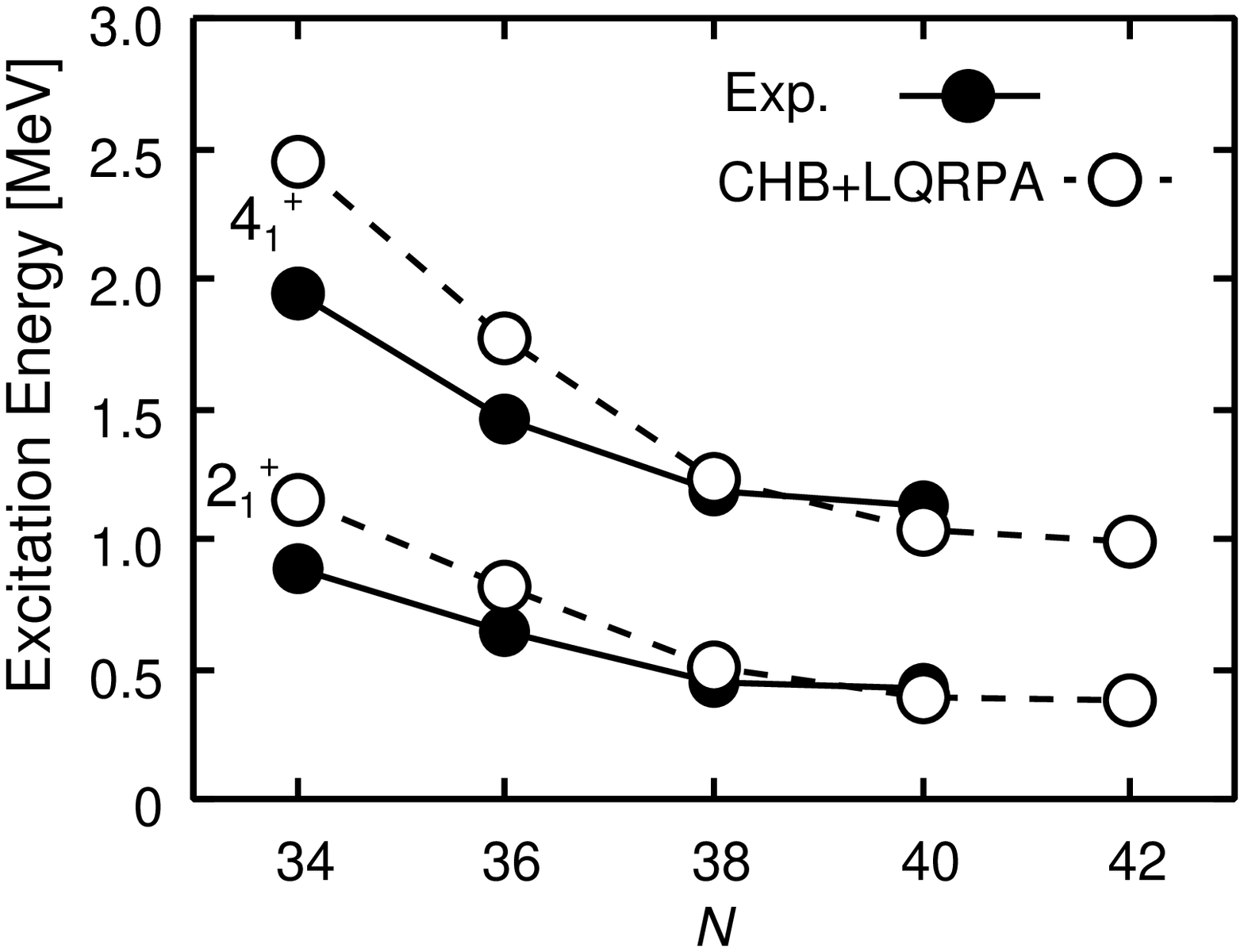}}
\subfigure[$R_{4/2}$]{
\includegraphics[height=0.3\textwidth,keepaspectratio,clip,trim=0 0 0 0]
{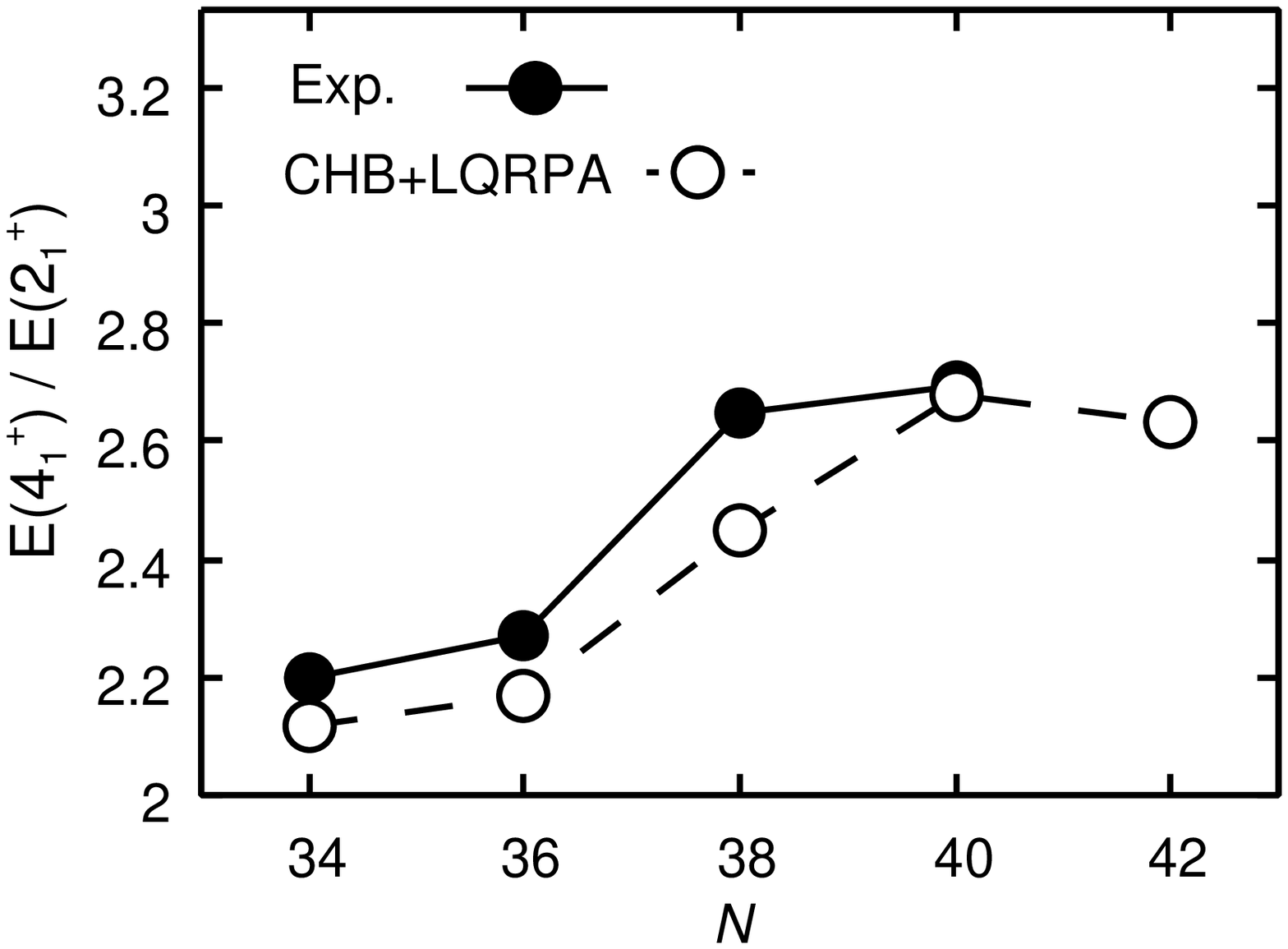}} 
\subfigure[$B(E2;2_1^+ \rightarrow 0_1^+)$]{
 \includegraphics[height=0.3\textwidth,keepaspectratio,clip,trim=0 0 0 0]
{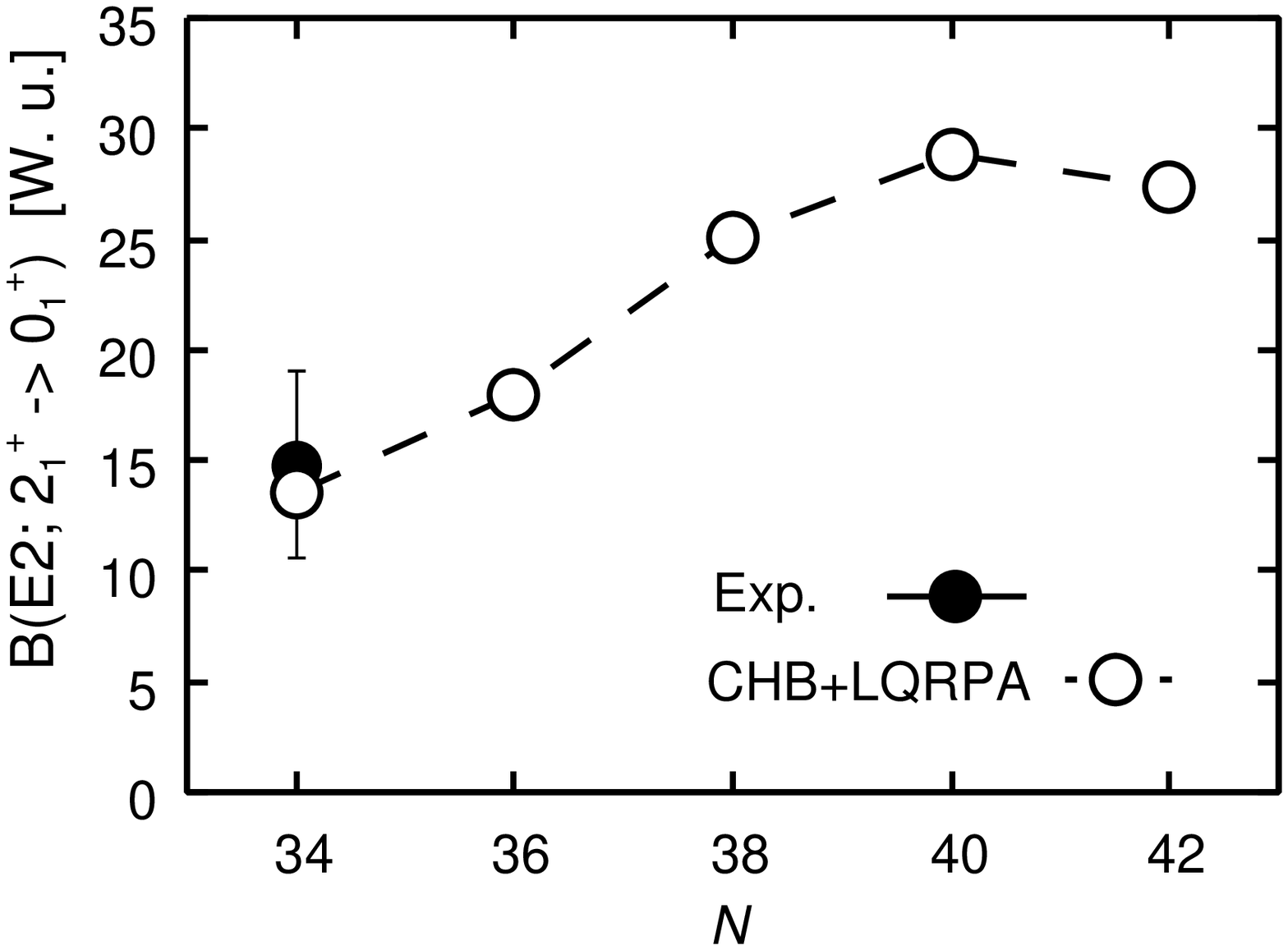}}
 \subfigure[$Q(2_1^+)$]{
\includegraphics[height=0.3\textwidth,keepaspectratio,clip,trim=0 0 0 0]
{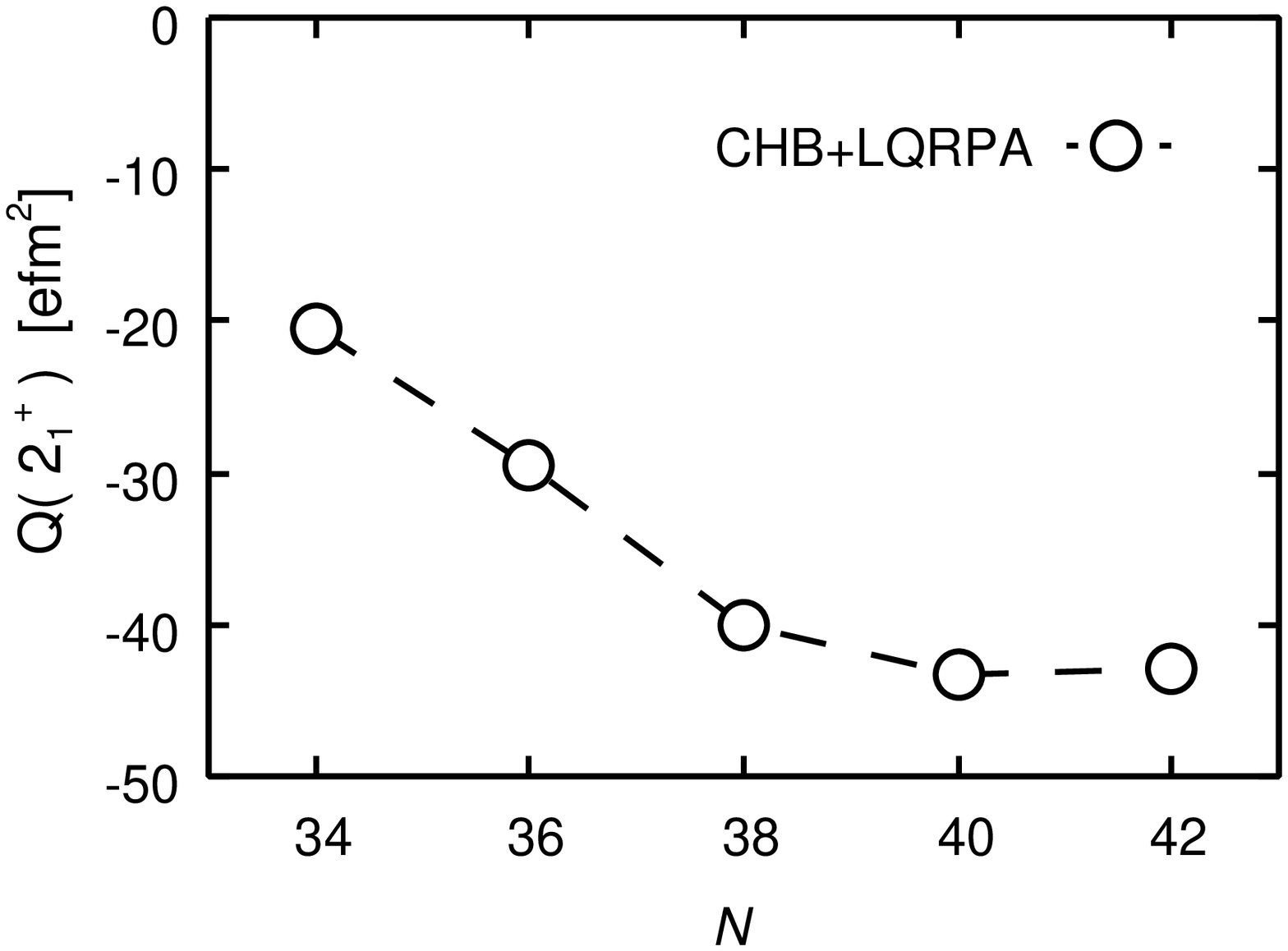}}
\end{center}
\caption{
Excitation energies of the $2_1^{+}$ and $4_1^{+}$ states,
their ratios $R_{4/2}$, $B(E2;2_1^+ \rightarrow 0_1^+)$,
and the spectroscopic quadrupole moments of the $2_1^{+}$ states
in comparison with the available experimental data \cite{Burger2005,Zhu2006,Aoi2009,Gade2010}.
The $B(E2)$ values are shown in Weisskopf units.
}
\label{fig:ERBQ}
\end{figure}

We show in Fig. \ref{fig:ERBQ} the calculated excitation energies of the $2_1^{+}$ and $4_1^{+}$ states,
their ratios $R_{4/2}=E(4_1^+)/E(2_1^+)$, $B(E2;2_1^+ \rightarrow 0_1^+)$,
and the spectroscopic quadrupole moments of the $2_1^{+}$ states $Q(2_1^+)$ 
calculated for $^{58-66}$Cr 
in comparison with the available experimental data.
Both of the experimental excitation energies and $R_{4/2}$ are reproduced well.
The decrease in the excitation energies and the increase in $R_{4/2}$
reflect the development of deformation with increasing the neutron number. 
The $E2$ transition and moments also indicate the enhancement of collectivity: 
the magnitude of $B(E2)$ and $Q(2_1^+)$ increase as the neutron number increases 
and they take a maximum at $N=40$. (Note that $Q(2_1^+)$'s are negative indicating prolate shapes.) 
To sum up, our result suggests the development of  prolate deformation 
from $N=34$ to $N=40$ and the largest collectivity at $N=40$.

%

\section{Concluding Remarks}
We have proposed the CHFB+LQRPA method for determining the inertial functions
in the 5D quadrupole collective Hamiltonian, with which one can take into
account the contributions from the time-odd components of the mean field to the
inertial functions.
We applied this method to 
the oblate-prolate shape coexistence in the low-lying states of \kr{72} 
and shape transition in neutron-rich Cr isotopes around $N=40$.
The calculated results are in good agreement with the available experimental data.

The CHFB+LQRPA method is based on the ASCC method 
and it can be used in conjunction with any interaction in principle. 
Nevertheless, in this study, we have employed a rather simple interaction, the P+Q model
including the quadrupole pairing interaction, for simplicity.
The implementation 
with 
modern energy density functionals, such as 
Skyrme energy functionals and density-dependent pairing interaction
is an issue for future under progress.

\ack
One of the authors (N. H.) is supported by the  
Special Postdoctoral Researcher Program of RIKEN.
The numerical calculations were carried out 
on SR16000 at Yukawa Institute for Theoretical Physics in Kyoto University
and 
RIKEN Cluster of Clusters (RICC) facility.
This work is supported by
KAKENHI (Nos. 21340073 and 20105003).

\section*{References}

\bibliographystyle{iopart-num}
\bibliography{text}

\end{document}